# Molecular Theory of Dendritic Liquid Crystals: Self Organization and Phase Transitions

Alexandros G. Vanakaras* and Demetri J. Photinos

*Department of Materials Science*
*University of Patras, Patras 26504, Greece.*
vanakara@upatras.gr    photinos@upatras.gr

**Abstract**

We formulate the configurational partition function for dendrimers, taking explicit account of their conformations and segmental interactions. Two approximate schemes are presented, one based on the effective dendrimer-dendrimer interactions and the other based on the interactions among the mesogenic units comprising the dendrimers. In the latter scheme, the description of the dendromesogenic system reduces to that of an ensemble of mesogenic dimers. Results of lattice calculations for phase transitions are presented showing that the minimal inclusion of shape anisotropy and of sub-molecular partitioning into chemically distinct parts reproduces the variety of phases and phase sequences observed experimentally and provides insights into the conformational aspect of these transitions.

**I. Introduction.**

The study of liquid crystalline phases formed by various types of dendritic systems [1-13] differing in their architecture, in the chemical structure of the branches, the functionalisation of the surface etc, has lead to the identification of several possible mechanisms of supermolecular self organisation[14]. These include:
**a.** Micro-segregation, generated by the partitioning of the dendritic structure into chemically distinct regions. This mechanism is believed to underlie the liquid crystalline self-organisation of dendrimers which lack orientability by virtue of their overall shape or of their subunits[1].
**b.** Mutual alignment of mesogenic units. This mechanism is dominant in radial or globular dendrimers containing mesogenic units. The latter could be attached to the external periphery of the dendritic scaffold [2-9] or be part of the scaffold as well [10]. The orientational order results primarily from the anisotropic interactions among mesogenic units belonging to different dendrimers (*inter*-dendritic) or to the same dendrimer (*intra*-dendritic). The intra-dendritic interactions could induce an asymmetry to the overall shape of the dendrimer that, in turn, enhances the orientational order. On the other hand, in systems with a flexible non-mesogenic dendritic scaffold and peripherally attached mesogenic units, the micro-segregation mechanism, stemming from the chemical distinction between the flexible scaffold and the mesogenic periphery, is superimposed to the mesogenic interactions and promotes partial positional order in the form of layering or column formation [2-9].

**c.** Self-assembly of dendritic units to form supramolecular structures that self-organise into liquid crystalline phases [11]. The structure and interactions of the dendra, which need not contain any intrinsically mesogenic segments, control the shape of the supramolecular entities and thereby the symmetry of the ordered phases.

**d.** Direct self-organisation of relatively rigid supermolecular structures, such as worm-like polymers or rigid rods [12,13] formed by the bonding of dendritic units. The shape of the superstructure is determined by the way in which the dendritic units are bonded while the overall rigidity can be controlled by the generation of the dendra.

A basic issue in formulating any molecular theory of dendritic mesomorphism is the extent of detail to which the structure, the conformations and the interactions of the dendritic units are to be described. The molecular size of these systems and the usually enormous number of conformational states they can access precludes a fully atomistic description and one has therefore to identify and retain only the elements that are of primary relevance to their mesomorphic behaviour. The complexity of the problem stems from the presence of an extensive hierarchy of interactions: dendrimers of a given generation have several topologically and chemically different segments and this gives rise to many combinations of *intra-* and *inter-*dendrimer segmental interactions. Here we consider two simplified views to this problem. One view is to assign distinct roles to intra- and inter-dendrimer interactions by treating the dendrimers as deformable objects [15-17] that can exist in a number of conformational states which are predetermined by the intra-dendrimer interactions. These objects are then assumed to interact with each other in a way that is dictated entirely by the *inter-*dendrimer segmental interactions. The other view is to consider directly the interactions among the dendritic segments in a pair-wise manner (i.e. ignoring 3-segment correlations or higher) and impose on the *intra-*dendritic pairs the configurational constraints dictated by their connectivity within the same dendrimer. In other words, this approach replaces the dendritic connectivity by a set of pair-wise configurational constraints on its segments.

The statistical mechanics approximations entailed by each of these two approaches are presented in section II. In section III we present calculations based on the deformable body approach for model dendrimers exhibiting interconverting calamitic-discotic states. The application of the segmental pair-interaction approach is illustrated in section IV for a model dendrimer consisting of a flexible non-mesogenic dendritic scaffold that is peripherally functionalised with mesogenic units. Finally, the conclusions are presented in Section V

**II. Statistical mechanics approximations**

We consider an ensemble of $N_D$ identical dendrimers labeled by the indices $I,J... = 1,2,3...N_D$. We denote the position of the $I^{th}$ dendrimer by $\mathbf{R}_I$, its orientation by $\Omega_I$ and the set of variables specifying its conformational state by $v_I$. The energy of the dendrimer at that state is $E(v_I)$. Let the interaction between two such dendrimers be described by the pair potential $U_{I,J} = U(\mathbf{R}_{I,J}; \Omega_{I,J}; v_I, v_J)$, with $\mathbf{R}_{I,J}, \Omega_{I,J}$ denoting respectively the position and orientation of dendrimer $J$ relative to $I$. The conformational energy $E(v_I)$ is understood to originate from the interactions among the segments that form dendrimer *I*. Labeling these segments by the index $i_I, j_I... = 1, 2, 3...N_S$, where $N_s$ is the total number of the segments into which the dendrimer is subdivided, and assuming that the intra-dendrimer interactions can be represented by a pair-wise superposition of interactions $u(i_I, j_I)$ among its constituent segments we have

$$E(v_I) = \sum_{i_I, j_I} u(i_I, j_I) \ . \tag{1}$$

Similarly, the dendrimer-dendrimer interaction potential $U_{I,J}$ is written as a superposition of interactions $u(i_I, j_J)$ among all the inter-dendrimer pairs of segments $i_I, j_J$,

$$U_{I,J} = \sum_{i_I, j_J} u(i_I, j_J) \ . \tag{2}$$

The equilibrium partition function for this ensemble of dendrimers is then

$$Q = \int d\{I\} \prod_{I=1}^{N_D} e^{-E^*(v_I)} \prod_{J=I+1}^{N_D} e^{-U^*_{I,J}} \tag{3}$$

with $\{I\}$ denoting collectively the complete set of configurational variables $\{\mathbf{R}_I; \Omega_I; v_I\}$ of the $N_D$ dendrimer ensemble, $E^*(v_I) = E(v_I)/k_B T$ and $U^*_{I,J} = U_{I,J}/k_B T$.

**IIa. Interconverting conformer formulation**
If we assume that the conformational states of the dendrimer are discreet, i.e. that the conformational variable assumes discreet values, then the formal integration in eq(3) will entail for each dendrimer in the system a summation over all conformations, $\sum_{v_I}$, and an integration, $\int d\varpi_I$, over its position and orientation variables denoted collectively by $\varpi_I = (\mathbf{R}_I; \Omega_I)$.

Suppose that the conformational states can be grouped into sets, with the states in each set exhibiting identical dendrimer-dendrimer interaction $U_{I,J}$. For example, in the special case where these interactions are assumed to be hard body repulsions, the grouping would be such that all the members of a set exhibit identical shapes for the dendrimer [16]. Thus, for brevity we will refer to these sets of conformations as "shapes", although the formulation is applicably to soft potentials as well. The different shapes of dendrimer *I* are denoted by $S_I$ and the distinct conformations associated with the same shape $S_I$ are denoted by $v(S_I)$. Then the conformational sum involving dendrimer *I* in eq (3) can be carried out first over all the conformations of a given set and then over all the sets, i.e.

$$\sum_{v_I} e^{-E^*(v_I)} e^{-U^*_{I,J}} = \sum_{S_I} [\sum_{v(S_I)} e^{-E^*(v_I)}] e^{-U^*_{I,J}} = \sum_{S_I} W_{S_I} G_{S_I, S_J}(\varpi_{I,J}) \ , \tag{4}$$

where the statistical weight $W_{S_I}$ of the state $S_I$ is defined by

$$W_{S_I} = \sum_{v(S_I)} e^{-E^*(v_I)} \ , \tag{5}$$

and

$$G_{S_I,S_J}(\varpi_{I,J}) = e^{-U^*_{I,J}} \qquad (6)$$

describes the interactions between any conformation of shape $S_I$ with any conformation of shape $S_J$, with $\varpi_{I,J}$ denoting the relative positional and orientational variables $\mathbf{R}_{I,J}, \Omega_{I,J}$ of the dendrimer pair *I, J*.

With this grouping, eq (3) can be put in the equivalent form

$$Q = \sum_{\{S_I\}} \int d\{\varpi_I\} \prod_{I=1}^{N_D} W_{S_I} \prod_{J=I+1}^{N_D} G_{S_I,S_J}(\varpi_{I,J}) \ . \qquad (7)$$

The configurational free energy of the system, $F = -k_B T \ln Q$, is approximated according to the variational cluster method [18] by introducing a variational weight function $\zeta_S(\varpi)$ for each set of *S* of conformations and retaining up to two-particle terms in the cumulant expansion. This leads to the following approximate expression for the free energy:

$$-F/N_D k_B T \approx \ln \zeta + \frac{1}{2}(N_D - 1)\ln \langle G \rangle \qquad (8)$$

where $\zeta = \sum_S W_S \zeta_S$, with $\zeta_S = \int d\varpi \zeta_S(\varpi)$ and the angular brackets denote averaging with respect to the probability distribution

$$\rho_S(\varpi) = W_S \zeta_S(\varpi)/\zeta \qquad (9)$$

namely,

$$\langle G \rangle \equiv \sum_{S_I,S_J} \rho_{S_I}(\varpi_I)\rho_{S_J}(\varpi_J) G_{S_I,S_J}(\varpi_{I,J}) \ . \qquad (10)$$

The variational functions are determined self consistently from the conditions

$$\zeta_S(\varpi) = \exp[\frac{\langle G_{S_I}(\varpi_I) \rangle - \langle G \rangle}{\langle G \rangle}] \ , \qquad (11)$$

with $\langle G_{S_I}(\varpi_I) \rangle = \sum_{S_J} \int d\varpi_J \rho_{S_J}(\varpi_J) G_{S_I,S_J}(\varpi_{I,J})$

The form of the dependence of the variational weight function $\zeta_S(\varpi)$ on the positional and orientational variables $\varpi_I = (\mathbf{R}_I; \Omega_I)$ reflects the symmetry of the phase. In our study of the mesomorphic behaviour of dendrimers we consider

- isotropic phases, for which $\zeta_S(\varpi)$ is independent of position and orientation,
- nematic phases, for which $\zeta_S(\varpi)$ is independent of position, $\zeta_S(\varpi)|_{Nem} = \zeta_S(\Omega)$,
- smectic phases, for which $\zeta_S(\varpi)$ is independent of the positional coordinates $X,Y$ in the plane of the smectic layers, $\zeta_S(\varpi)|_{Sm} = \zeta_S(Z;\Omega)$, *Z* being the positional coordinate along the layer normal,
- columnar phases, for which $\zeta_S(\varpi)$ is independent of the positional coordinate *Z* along the columnar axis, $\zeta_S(\varpi)|_{Col} = \zeta_S(X,Y;\Omega)$, where the coordinates $X,Y$ define the plane normal to the columnar axis of the phase,

- plastic crystal phases, for which $\zeta_S(\varpi)$ is independent of the orientational variables, $\zeta_S(\varpi)|_{PCr} = \zeta_S(X,Y,Z)$, and
- crystal phases, for which $\zeta_S(\varpi)$ has the full dependence on position and orientation, $\zeta_S(\varpi)|_{Cr} = \zeta_S(X,Y,Z;\Omega)$.

For the inter-converting conformer approach [15-17] to be usable in practice it is necessary that the number of the relevant shapes be not too large so that a reasonable number of intrinsic weights $W_S$ and interactions $G_{S_I,S_J}(\varpi_{I,J})$ be required as input to the calculation. These quantities can be furnished to some coarse grained representation by considering first the conformation statistics and segmental interactions of a single dendrimer in isolation. The basic computational step is then the solution of the self-consistency equations, following which the probability distribution of eq(9) and the corresponding free energy of eq (8) are obtained. The pair distribution function in this approximation is given by $\rho^{(2)}_{S_I,S_J}(\varpi_I,\varpi_J) = \rho_{S_I}(\varpi_I)\rho_{S_J}(\varpi_J)G_{S_I,S_J}(\varpi_{I,J})/\langle G \rangle$.

The important conformation sets, or shapes, to keep in eq(9) are not necessarily the ones that have the highest intrinsic probabilities $P_S^0 = W_S / \sum_{S'} W_{S'}$ for an isolated dendrimer but rather the ones that have significant probabilities to occur in the bulk phase of the interacting dendrimers. The probability for finding a conformational state of shape $S$ is given by $P_S = W_S \zeta_S / \zeta$, from which it is seen that for a shape to have significant probability in the bulk phase it is necessary that, aside from its intrinsic weight, it should interact with its environment favourably enough to acquire a large value of its self-consistent weight function integral $\zeta_S$.

As shown in section III, model calculations employing just two basic shapes and very simple forms for their interactions can reproduce a rich variety of phase transitions and the associated structural and conformational changes. The whole approach, however, becomes clearly inadequate for the description of phenomena where the internal motions of the dendrimer are directly involved. Thus, for example, the static dielectric properties can be described by assigning to each shape a molecular polarisability tensor and a molecular dipole moment (usually vanishing by symmetry) and considering reorientations and translations of the dendrimer as a whole. This constitutes, of course, a severely oversimplified molecular picture of the dielectric behaviour of the actual systems, except perhaps for extreme cases of internally very dense dendrimers precluding any significant intra-dendrimer segmental rearrangements[16]. The picture can be improved by considering distributed polarisabilities and dipole moments over the dendrimer volume and by further allowing for some deformation modes of the basic shapes. This however makes the approach more complex and requires a larger set of input information to be furnished with the guidance of the atomistic structure and conformational statistics of the single dendrimer. Similarly, the description of the dynamic behaviour of the dendrimers with this approach is meaningfull only for time scales pertaining to global motions of the dendrimer shapes and of their basic deformation modes while it is clearly inapplicable for time scales associated with intra-dendrimer segmental rearrangements.

**IIb. Segmental formulation.**

Rather than viewing the entire dendrimer as the basic molecular unit, this description focuses on the sub-dendritic units that give rise to the mesomorphic behaviour of the dendrimer ensemble. Denoting the segmental interaction Boltzmann factors by $G(i_I, j_J) = \exp(-u(i_I, j_J)/k_B T)$, we may rewrite eq(3), taking into account eqs(1), (2), as follows

$$Q = \int d\{I\} \prod_{I=1}^{N_D} \prod_{i_I \neq j_I} G(i_I, j_I) \prod_{J=I+1}^{N_D} \prod_{i'_I, j'_J} G(i'_I, j'_J) \ , \tag{12}$$

in which the interaction terms associated with *intra*-dendrimer pairs of segments are explicitly separated from those of the *inter*-dendrimer pairs. In a more compact notation, all the segments in the system are labeled by a single index $i, j... = 1, 2, 3... N_S \times N_D$, without specific reference to the dendrimer they belong. The partition function expression assumes then the compact form

$$Q = \int d\{I\} \prod_{i \neq j} G(i, j) \ , \tag{13}$$

with the understanding that the different pairs *i, j* are not necessarily equivalent.

The variational cluster method can then be applied to eq(12) to obtain the free energy of the system up to two-segment terms in the cumulant expansion. Denoting the variational weight function introduced for each segment *i* by $\zeta(i)$, we have for the probability distribution of a single segment

$$\rho(i) = \zeta(i)/\zeta_i \tag{14}$$

with $\zeta_i \equiv \int di \zeta(i)$ denoting the integrated weight function for segment *i*. The probability distribution for a pair of segments is given in this approximation by

$$\rho^{(2)}(i, j) = \rho(i)\rho(j)G(i, j)/\langle G_{i,j} \rangle \ , \tag{15}$$

where

$$\langle G_{i,j} \rangle \equiv \int d(i) d(j) \rho(i) \rho(j) G(i, j) \ , \tag{16}$$

and the expression for the approximate free energy is

$$-F/k_B T \approx \sum_i \ln \zeta_i + \sum_{i \neq j} \ln \langle G_{i,j} \rangle \ . \tag{17}$$

The variational weight functions are determined by functional minimization of the above expression for the free energy which leads to the self-consistency equations

$$\zeta(i) = \exp\left[ \sum_j \frac{\langle G_j(i) \rangle - \langle G_{i,j} \rangle}{\langle G_{i,j} \rangle} \right] \ , \tag{18}$$

where

$$\langle G_j(i) \rangle = \int d(j) \rho(j) G(i, j) \ . \tag{19}$$

The dependence of the variational weight function $\zeta(i)$ on the positional and orientational variables reflects the symmetry of the phase under consideration, as detailed in the previous section, only now these variables refer to the $i^{th}$ segment rather that the entire dendrimer.

The segmental approach rests on the partitioning of the conformational energy of the dendrimer among the *intra*-dendritic pairs of segments, as indicated in eq ((12)). This cannot always be done in a unique way. However, rather than addressing this issue in a general way, we give in section IV a specific example of the application of this approach to a globular dendrimer functionalized peripherally with mesogenic units. In any case, the input information for the calculations with the segmental approach are the interaction potentials for the pairs of *inter*-dendritic segments as well as those of the *intra*-dendritic pairs, the latter differing from the former in that they include the additional energetic contribution associated with their connectivity through the dendritic scaffold.

**III. Lattice calculations for an inter-converting rod-plate model of dendromesogens**

The inter-converting conformer formulation of section IIa will be applied here to a very primitive model: the dendrimer is assumed to exhibit just two sets of conformations, or "shapes", a rod-like and a disc-like, for which the shape indices *S=r* and *S=d* will be used. In accordance with this crude molecular picture, the interactions are modelled as purely repulsive, with their strength being determined by the extent of overlap between the molecular volumes. Molecular partitioning, and the associated microsegregation of the self-organisation, is introduced into the model by differentiating between the strength of the repulsive interactions among different parts of the rod-like or disc-like objects. In keeping with the simplicity of this modelling, the calculations of the free energy and the self-consistent weight functions of eqs.(8), (11) are performed on a cubic lattice. Thus, we have assumed that the molecules are made up of cubical blocks and are constrained to translate and rotate on a cubic lattice space. The lattice unit cell dimensions are taken to coincide with the size of the molecular building blocks. By restricting the molecular conformers to move so that each of their building blocs occupies a single unit cell of the lattice, the computational effort is reduced considerably, compared to a continuous sampling of the molecular positions and orientations, without severely affecting the predictive aspects of the model.

The same building blocks are used to construct both the rod and the disc conformers of the dendrimer. The blocks of dendrimer *I* are enumerated by the index $b_I$. As shown in figure 1, two types of blocks are introduced in order to differentiate between chemically distinct parts of the dendrimer. In the simple instance of figure 1, the differentiation is between regions with high density of mesogenic units (dark grey blocks) and regions populated by the flexible chain segments that form the dendritic scaffold (light grey blocks).

The molecular interactions used for the present lattice calculations are modeled in an pair-wise additive scheme such that the potential between a pair of blocks belonging to different dendrimers vanishes except when these blocks occupy the same or adjacent lattice sites. Accordingly the interaction potential between molecules *I* and *J* can be written as

$$U_{I,J} = \sum_{b_I,b_J} \left( u^{(0)}_{b_I,b_J} \delta(r_{b_I,b_J}) + u^{(1)}_{b_I,b_J} \delta(|r_{b_I,b_J}|-1) \right) , \qquad (20)$$

where $r_{b_I,b_J}$ denotes the distance between blocks $b_I$ and $b_J$, and $u^{(0)}_{b_I,b_J}$, $u^{(1)}_{b_I,b_J}$ stand for the values of the potential for that pair of blocks occupying respectively the same lattice site or adjacent lattice sites.

To study how molecular partitioning influences the phase behaviour of these systems we have considered two different models of the block-block interactions. In the first parameterisation we take the two kinds of sub-molecular blocks to have identical hard-body repulsions, i.e. two blocks, of any kind, cannot occupy the same lattice site and have otherwise no interaction. In the second parameterisation, the following differences are introduced between the flexible scaffold blocks and the mesogenic blocks:
(i) a scaffold block cannot occupy the same lattice site with another block, of any kind, has no interaction with adjacent scaffold blocks and exerts a soft repulsion to any mesogenic block occupying an adjacent site, and
(ii) a mesogenic block exerts soft repulsions, of different intensity, to other mesogenic blocks occupying the same or adjacent lattice sites.
The parameters associated with eq. (20) for the above two models of block-block interactions are summarised in table 1.

| $b_1$ \ $b_2$ | Scaffold | Mesogens |
|---|---|---|
| Scaffold | ∞ (0) | ∞ ($u_b/2$) |
| Mesogens | ∞ ($u_b/2$) | $u_b$ ($u_b/4$) |

**TableI.** Interaction parameters between the building blocs when they occupy the same lattice site and when they are in adjacent lattice sites (values in bracets). For the calculations we have used $u_b/k_BT = 0.1$.

Having specified the intermolecular interactions, we proceed with the calculation of the phase diagrams for the inter-converting rod-disc model of the dendromesogens. We have solved the self consistency conditions of eq. (11) and have calculated phase equilibrium, based on the free energy of eq. (8), for isotropic, nematic, orthogonal smectic and rectangular orthogonal columnar phases. The calculation is initiated by giving a specific value for intrinsic probability of the disclike molecular state $P_d^0$. Then, the dimensionless pressure $p^* = pV_{mol}/k_BT$ (here $p$ is the actual pressure and $V_{mol}$ the molecular volume), at which a phase transition occurs, is located by solving the coexistence conditions for the two phases. The resulting phase diagrams for the two parameterizations of the model interaction are presented in figures 2(a),(b). In both cases the intrinsic probability of the disclike molecular state, $P_d^0$, is varied in the range from 0 (purely rodlike conformers) to 1 (only disclike conformers).

Although the phase diagrams of figures 2(a),(b) posses all the essential information for the phase stability of the system, it is experimentally more relevant to introduce the temperature as the thermodynamic variable instead of the intrinsic probability of the molecular conformations. To do that we assume that the intrinsic weights for the rod-like (disc-like) conformers may be written as $W_{r(d)} = \exp[-\varepsilon_{r(d)}/k_BT]$, with $\varepsilon_{r(d)}$ representing effective free energies for the two conformers. On further assuming that the free energy difference $\varepsilon_r - \varepsilon_d$ does not vary with temperature appreciably (although the individual free energies $\varepsilon_r$ and $\varepsilon_d$ may do so), we obtain for the scaled reciprocal temperature $(\varepsilon_r - \varepsilon_d)/k_B T = \ln\left(P_d^0/(1-P_d^0)\right)$. Based on this expression, the $(p^*, P_d^0)$ phase diagram of figures 2(a),(b) are readily transformed into the $(pv_0/\Delta\varepsilon, \Delta\varepsilon/kT)$ phase diagrams shown in figures 3(a)-(d).

**Phase behaviour of the model systems**
As evident from the phase diagrams in figures 3(a),(b), when the intermolecular interactions do not distinguish between different parts of the molecules, as in the case of the purely hard body repulsions, the ensuing phase diagrams show the usual phase sequences obtained in hard-rod or hard-disc molecular theories [17,19] and simulations [20,21]. In particular, when the rod like conformers have larger intrinsic weight than the disc like ($\varepsilon_r < \varepsilon_d$), the system transforms on cooling from the isotropic phase to a uniaxial nematic phase and from there to a positionally ordered phase. In the case of the purely hard repulsive system the positionally ordered phase is columnar, formed by rods that are free to slide side-by-side along the columnar axis of the phase. A similar disappearance of smectic phases in favour of the columnar self-organization has been reported [22,23] for highly oriented systems of hard cylindrical objects. For relatively low pressures, the calamitic nematic phase is suppressed and the only phase transition is from the isotropic to the columnar phase at rather low temperatures. On the other hand, when the disclike conformer is the one with the larger intrinsic weight ($\varepsilon_d < \varepsilon_r$), the phase sequence, at all the pressures, goes, on cooling, from isotropic to a discotic nematic phase and then on to a columnar phase.

The situation changes dramatically when the submolecular partitioning is incorporated, even in the minimal and perhaps oversimplified way it is done in the present calculations. Indeed, on introducing differentiating interactions between the two species of sub-molecular blocks, the variety of phases and of the possible phase sequences becomes much richer, as it is evident from the diagrams in figures 3(c),(d). More specifically, when the intrinsically more abundant conformer is the rod-like, five phases of different symmetries appear on the pressure vs temperature phase diagram namely, the isotropic, two uniaxial nematic phases (one calamitic and one discotic denoted by $N_r$ and $N_d$ respectively), the orthogonal smectic phase (rich on rodlike conformers) and the columnar discotic phase (rich in disc-like conformers). At low pressures the phase sequence is similar to that of the purely repulsive system, i.e. isotropic / discotic nematic /columnar. At moderate pressures, or, equivalently for lower free energy difference of the molecular states, the orthogonal smectic phase is inserted between the discotic nematic phase and the columnar phase. This phase transformation, from lamellar to columnar, is rarely observed in common, low molar mass liquid crystals and, in fact, the few known instances [24] involve some self-assembly of the entities that self-organise into columns. There are, however, at least two cases of such lamelar-columnar phase transitions reported in the literature. J.-M. Rueff *et al* [8] and R. M. Richardson *et al* [5], in both of which the underlying mechanism is related to the change in the dominant conformation of the dendrimers, in accordance with the results of the present calculations.

When $\varepsilon_d - \varepsilon_r \ll 1$ but still with $\varepsilon_d < \varepsilon_r$, corresponding to nearly equal intrinsic weights of the rod-like and the disc-like conformers, a phase sequence becomes possible whereby, on cooling from the isotropic, a discotic nematic phase is obtained which, on further cooling, is transformed to a rod (calamitic) nematic which in turn cools to an orthogonal smectic phase. This result shows that the intrinsically more abundant conformer in not necessarily the dominant one in the ordered bulk phase. This is demonstrated in figure. 4, where we have plotted the calculated bulk probabilities $P_r$ and $P_d$ of the two conformers as function of pressure, at scaled temperature $(\varepsilon_r - \varepsilon_d)/k_B T = 0.71$. The plots indicate that all the phase transitions are accompanied by significant chances of the bulk probabilities of the two conformers. However the transition from smectic to columnar is accompanied by a clear and abrupt inversion of the conformer populations.

**IV. The dimer ensemble model of dendromesogens.**

In this section we present a concrete application of the segmental approach outlined in section II. To this end we consider the dendritic architecture shown in figure 5. The dendrimer has a flexible scaffold of symmetrical radial topology, with three branches radiating at each branching point. The periphery of the dendrimer is functionalised by attaching identical rod-like mesogenic units to the terminal branches. This architecture is representative of a class of extensively studied dendromesogens [2,5].

Two mesogenic units $i, j$ belonging to different dendrimers are said to form a non-bonded pair, such as the pair AB' in figure 5, and are assigned an interaction $u_n(i,j) = u_n(\mathbf{r}_{ij}, \omega_{ij})$ that depends on the relative position $\mathbf{r}_{ij}$ and the relative orientation $\omega_{ij}$ of these units. The interaction of bonded pairs, i.e pairs of mesogenic units belonging to the same dendrimer, consists of the non-bonded interaction and an additional term $u_{path}$ associated with the conformational constraints imposed on the relative positions and orientations of the two units as a result of their attachment at the ends of a path formed by chain segments (branches) of the scaffold. Accordingly, the bonded potential $u_b$ for a pair $i,j$ is a function of the conformational variables $v_{i,j}$ of the branch-path that connects them,

$$u_b(i,j) = u_b\{v_{ij}\} = u_n(\mathbf{r}_{i,j}\{v_{ij}\}, \omega_{i,j}\{v_{ij}\}) + u_{sp}\{v_{ij}\} \qquad (21)$$

For simplicity, the flexible branches of the scaffold are not directly assigned any anisotropic interaction. They are assumed to exist in a number of discreet conformations, generated according to the RIS scheme [25] with equal intrinsic probabilities, and aside from that they simply fill space in the dendritic interior.

Under these assumptions, the pair-wise sum in the free energy expression of eq (1.17) has five distinct types of pair terms $\langle G_{i,j} \rangle$ for the specific example of fourth generation dendrimer shown in figure 5. One type is formed by mesogenic units belonging to different dendrimers i.e. non-bonded pairs *AB'* in the figure 5. With the $N_S = 24$ mesogenic units of each dendrimer being identical, and equivalent, by virtue of their symmetric attachment on the dendritic periphery, there are altogether $NN_S(N_D - 1)/2$ equivalent non-bonded pair terms $\langle G_{i,j} \rangle \equiv \langle G_n \rangle$ in an ensemble of $N_D$ dimers (which therefore contains in total $N = N_S N_D$ mesogenic units).

The other four types of pair terms $\langle G_{i,j} \rangle$ are *intra*-dendritic, formed, as shown in figure 5, by first ($AB_1$) second ($AB_2$) third ($AB_3$) and fourth ($AB_4$) neighbours on the same dendrimer. We shall denote these bonded-pair terms by $\langle G_b \rangle$, with the index $b = 1, 2, 3, 4$ specifying their neighbour order within the dendrimer. With the number of equivalent pairs of neighbour order $b$ denoted by $Nh_b$ in the ensemble, we have for the factors the values $h_1 = 1/2, h_2 = 1, h_3 = 2, h_4 = 8$.

According to the above considerations, and noting that the *N* mesogenic units in the system, being equivalent, are described by a single integrated weight $\zeta$, the segmental summation in the free energy of eq (1.17) can be carried out to yield the expression

$$-F/Nk_BT = \ln\zeta + \frac{1}{2}N_S(N_D-1)\ln<G_n> + \sum_{b=1}^{4}h_b\ln<G_b> \qquad (22)$$

The bonded contributions of different neighbour orders in this expression correspond to typical mesogenic dimers, i.e pairs of mesogenic units linked by a flexible spacer[26]. The spacer in this case is the branch-path that connects the mesogenic pair. Accordingly, the dendrimer system is treated as an ensemble of mesogenic dimers with identical terminal units but different spacers. The proportion of each type of spacer in the dimer ensemble is governed by the factors $h_b$. From this stage on, the results obtained within the molecular theory of liquid crystalline dimers [27-29] can be carried over straightforwardly to the description of the dendrimers, providing a way to phase transition calculations, evaluation of segmental order parameters, pair correlation averages and dynamic response on the time scale associated with segmental motions. These calculations start from the interactions of the mesogenic units and the conformation structure of the pertinent branch-paths within the dendritic scaffold[30]. It is known that the mesomorphic behaviour of dimers could be very sensitive to the structure and length of the spacer [26]. This sensitivity is carried over to the individual bonded-pair terms that are superimposed to form the free energy of eq(1.22).

## V. Conclusions

We used the variational cluster method to approximate, to second order terms in the cumulant expansion, the free energy of an ensemble of dendrimers. We developed two approximation schemes, a global and a segmental one, in order to address different aspects of the molecular theory of liquid crystalline dendrimers.

In the global approach the dentrimers are viewed as the elementary entities in the ensemble. This approach is therefore suitable for the description of the self organisation of the entire dendrimers in terms of their dominant conformations and the respective global dendrimer-dendrimer interactions. These are dictated by a coarse grained parameterisation of the atomistic structure of the dendrimer. Even in its most primitive form (with dendrimers consisting of two chemically distinct components and restricted to just two uniaxial conformations that are allowed to move on a cubical lattice) the global approach accounts for all the mesophases that can be obtained from uniaxial objects. It also predicts a rich variety of phase sequences depending on the intrinsic populations of the two conformations. In particular, it accounts for the lamellar-columnar phase transitions and predicts nematic-nematic transitions associated with changes in the dominant conformations of the dendrimer.

In the segmental approach the basic entities in the ensemble are identified with the mesogenic groups of the dendrimer. The connectivity of these groups within the dendrimer is conveyed by an effective potential that can be derived from the conformational statistics of the dendritic scaffold. With this approach, the ensemble of dendromesogens reduces to an ensemble of mesogenic dimers with spacers of different lengths, corresponding to the different branch paths within the dendritic scaffold. The segmental approach is suitable for the description of the mesomorphic properties that are sensitive to the ordering and the motion of dendritic segments rather than the dendrimer as a whole. However, it should be kept in mind that dimer picture of the segmental approach is obtained by ignoring correlations of more than two mesogenic segments at a time, and therefore its reliability is restricted to situations not involving strongly correlated collective movements of the dendrimer constituents.

**Acknowledgments**

Support from the RTN Project "Supermolecular Liquid Crystal Dendrimers – LCDD" (HPRN-CT2000-00016) is acknowledged. AGV acknowledges support through the "Caratheodores" research programme of the University of Patras.

**References**

[1]   U. Stebani, G. Lattermann, M. Wittenberg and J.H. Wendorff, *J. Mater. Chem.*, 1997, **7**, 607.
[2]   J. Barberá, M. Marcos and J. L. Serrano, *Chem. Eur. J.,* 1999, **5**, 1834.
[3]   G. H. Mehl and J. W. Goodby, *Chem. Ber.*, 1996, **129**, 521.
[4]   G. H. Mehl and J. W. Goodby, *Angew. Chem. Int. Ed. Engl.*, 1996, **35**, 2641; R. Elsäβer, G. H. Mehl, J. W. Goodby and D. J. Photinos, *Chem. Commun.*, 2000, **10**, 851.
[5]   R. M. Richardson, S. A. Ponomarenko, N. I. Boiko and V. P. Shibaev, *Liquid Crystals,* 1999, **26**, 101.
[6]   M. Marcos, R. Giménez, J. L. Serrano, B. Donnio, B. Heinrich and D. Guillon, *Chem. Eur. J.*, 2001, **7**, 1006.
[7]   B. Donnio, J. Barberá, R. Giménez, D. Guillon, M. Marcos and J. L. Serrano, *Macromolecules,* 2002, **35**, 370.
[8]   J.-M. Rueff, J. Barbera, B. Donnio, D. Guillon, M. Marcos, and J.-L. Serrano, *Macromolecules*, 2003, **36**, 8368.
[9]   V. Percec, P. Chu, G. Ungar and J. Zhou, *J. Am. Chem. Soc.,* 1995, **117**, 11441.
[10]  L. Gehringer, D. Guillon and B. Donnio, Macromilecules, 2003, **36**, 5593.
[11]  V. Percec, W.-D. Cho and G. Ungar, *J. Am. Chem. Soc.,* 2000, **122**, 10273. and references therein.
[12]  A.D. Schluter and J.P. Rabe, *Angew. Chem., Int. Ed.,* 2000, **39**, 864.
[13]  N. Ouali, S. Mery, and A. Skoulios, *Macromolecules*, 2000, **33**, 6185.
[14]  C. Tschierske, *J. Mater. Chem.*, 2001, **11**, 2647.
[15]  A.F. Terzis, A.G. Vanakaras and D.J. Photinos, *Mol. Cryst. Liq. Cryst.,* 1999, **330**, 517.
[16]  A.F. Terzis, A.G. Vanakaras and D.J. Photinos, *Mol. Cryst. Liq. Cryst.,* 2000, **352**, 265.
[17]  A.G. Vanakaras and D.J. Photinos, *J. Mater. Chem.,* 2001, **11**, 2832.
[18]  T. Tanaka, *Methods of Statistical Physics*, (Cambridge: University Press) (2002).
[19]  A. G. Vanakaras and D. J. Photinos, *Molec. Cryst. Liq. Cryst.*, 1997, **299**, 65.
[20]  S. C. McGrother SC, D. C. Williamson and G. Jackson, *J. Chem. Phys.*, 1996, **104**, 6755.
[21]  G. A. C. Veerman and D. Frenkel, *Phys. Rev. A*, 1992, **45**, 5632.
[22]  A. Stroobants, H. N. W. Lekkerkerker and D. Frenkel, *Phys. Rev. A*, 1987, **36**, 2929.
[23]  A. M. Somoza and P. Tarazona, *Phys. Rev. A*, 1989, **40**, 4161.
[24]  S. Diele, S. Grande, J. Kain, G. Pelzl and W. Weissflog, *Mol. Cryst. Liq. Cryst.*, 2001, **362**, 111.
[25]  P. J. Flory, "Statistical mechanics of chain molecules", Interscience Publishers, (1969).
[26]  C. T. Imrie and G. R. Luckhurst, Chapter X, In *Handbook of Liquid Crystals Vol. 2B, Low Molecular Weight Liquid Crystals*, edited by D. Demus, J. Goodby, G. W Gray, H. –W. Spiess, V. Vill, (Wiley-VCH ), (1998).
[27]  P. K. Karahaliou, A. G. Vanakaras and D. J. Photinos, *submitted.*
[28]  H. S. Serpi and D. J. Photinos, *J. Chem. Phys*, 1996, **105**, 1718.
[29]  H. S. Serpi and D. J. Photinos, *Mol. Cryst. Liq Cryst.*, 2000, **352**, 205.
[30]  A. G. Vanakaras and D. J. Photinos, *in preparation.*

# Figure Captions

**Fig.1** Schematic representation of the two dominant conformers (rod- and disc-like) of a globular dendrimer peripherally functionalized with mesogenic units, together with their model structures used in the lattice calculations.

**Figure 2.** Calculated ($p^*$, $P_d^0$) phase diagrams (dimensionless pressure vs intrinsic probability of the disclike conformer) for the interconverting rod-disc model of the dendromesogens for the two parameterizations of the block-block interactions: (a) purely hard-body repulsions between the blocks and (b) selective interactions between blocks corresponding to the mesogenic units and to the dendritic scaffold.

**Figure 3.** Calculated phase diagrams (pressure vs reciprocal temperature) for the inter-converting rod-disc model of the dendromesogens for the two parameterizations of the block-block interactions: (a, b) purely hard-body repulsions between the blocks with the rod-like conformer of lower intrinsic free energy than the disc-like (a) and vice versa (b). (c,d) the corresponding graphs for the segment-differentiating parameterization shown in table I.

**Figure 4.** Calculated bulk probability of the rodlike and disclike conformers as a function of pressure for the system whose phase diagram is given in figure 4(d) at the fixed value of the scaled temperature $(\varepsilon_r - \varepsilon_d)/k_B T = 0.71$.

**Figure 5.** Fourth generation dendrimer topology showing the hierarchy of interactions among *intra*-dendrimer mesogenic pairs $AB_b$ according to the order of the branching points *b=1,2,3,4* present in the branch path connecting the members of each type of pair. An *inter*-dendrimer pair *AB'* is also indicated.

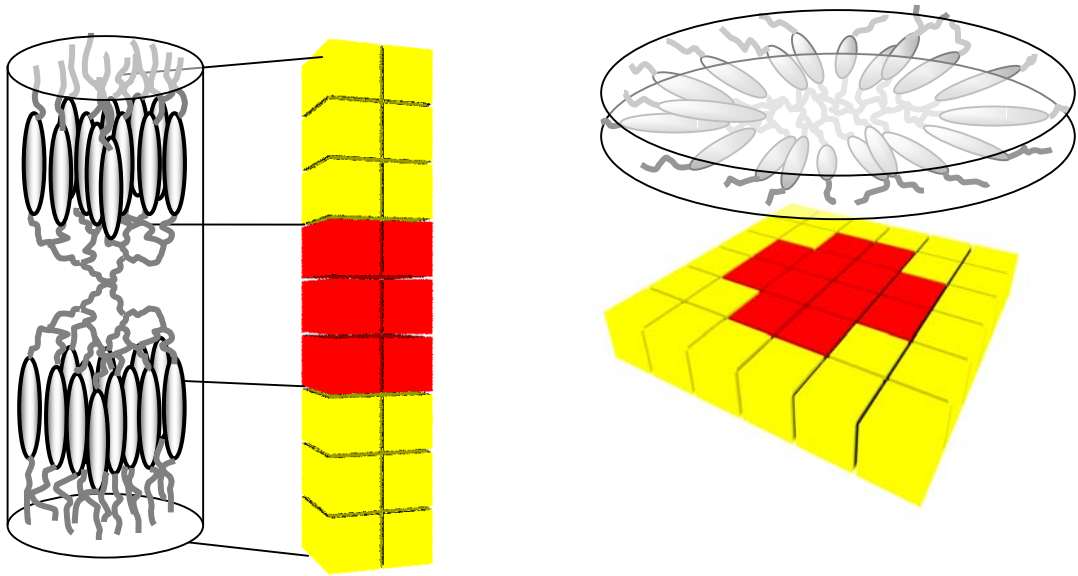

FIGURE 1

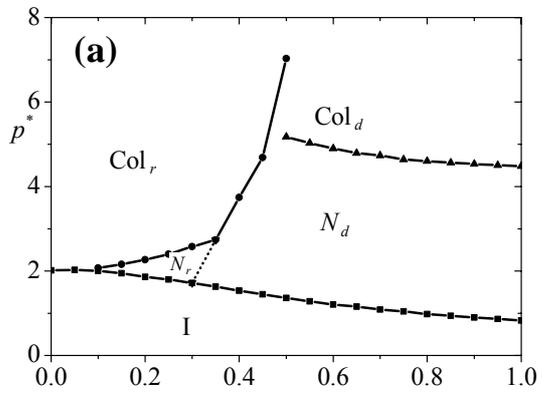 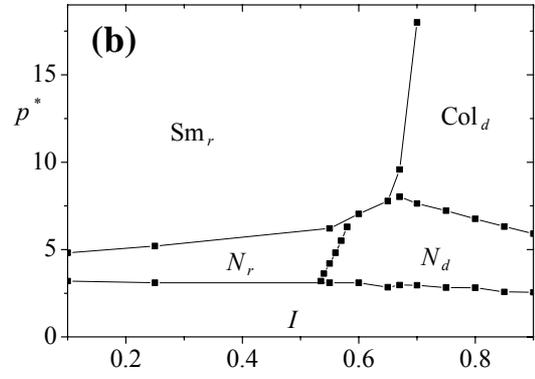

FIGURE 2

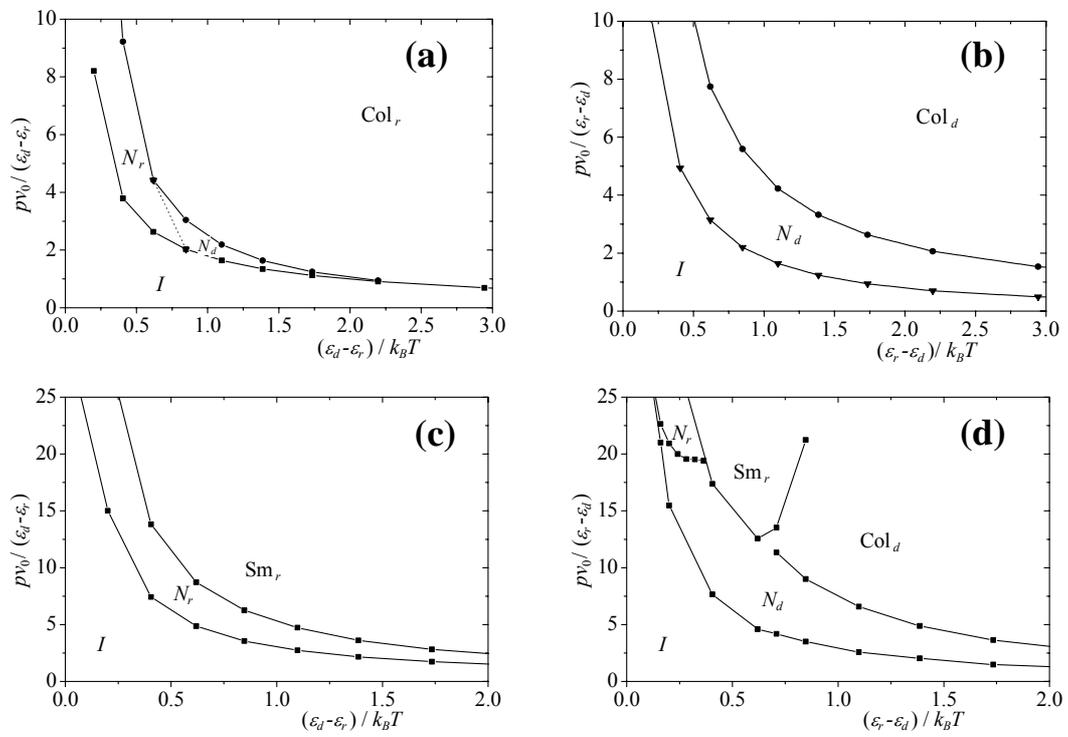

FIGURE 3

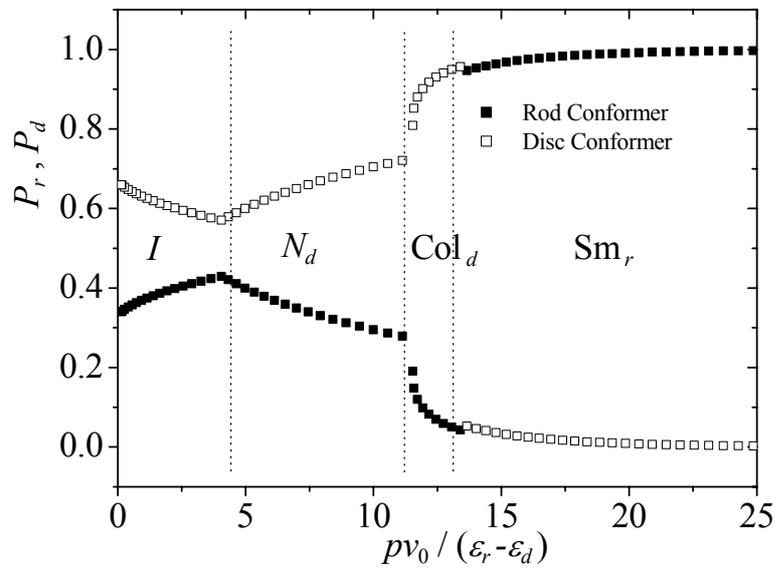

FIGURE 4

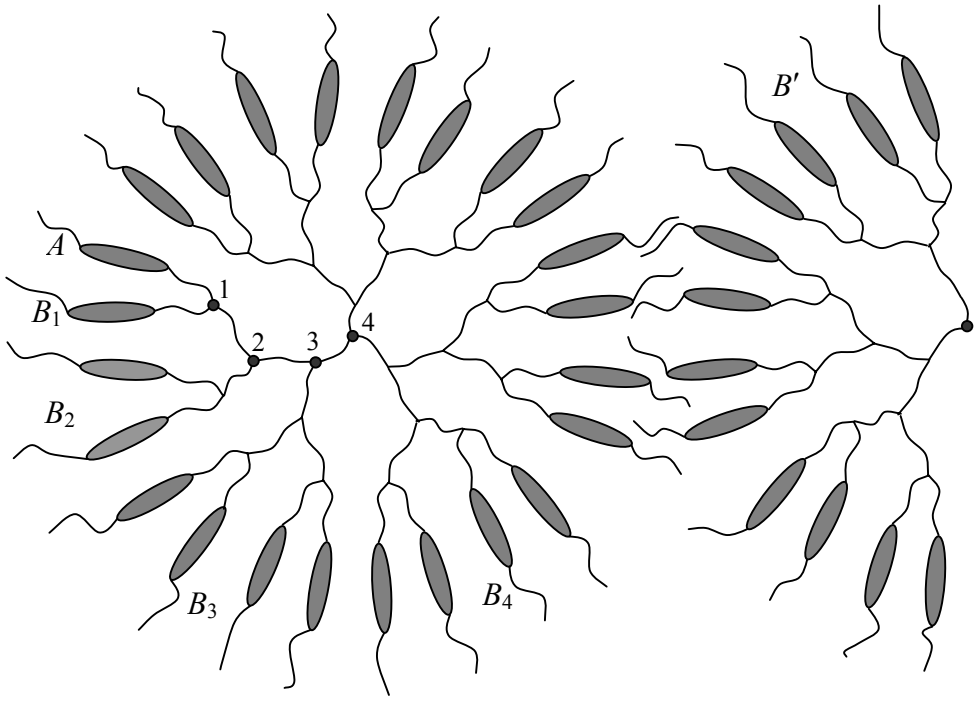

FIGURE 5